\documentclass[twocolumn]{aastex702}

\newcommand{\Aani}{A_{\rm ani}}

\shorttitle{Mock Validation of a Density Diagnostic}
\shortauthors{Kutlimuratov \& Tadjibayev}

\begin{document}

\title{Power-law Coupling and Mock Validation of a Surface-density Anisotropy Diagnostic for 85 Rich Galaxy Clusters}

\author[orcid=0000-0001-8681-9137,gname=Sardor Sharipbayevich,sname=Kutlimuratov]{Sardor Sharipbayevich Kutlimuratov}
\affiliation{Department of Physics, Chirchik State Pedagogical University, 104 Amir Temur Street, Chirchik 111700, Uzbekistan}
\email[show]{s.kutlimuratov@cspu.uz}
\correspondingauthor{Sardor Sharipbayevich Kutlimuratov}

\author[gname=Ikram Uralbayevich,sname=Tadjibayev]{Ikram Uralbayevich Tadjibayev}
\affiliation{Department of Physics, Chirchik State Pedagogical University, 104 Amir Temur Street, Chirchik 111700, Uzbekistan}
\email[show]{ikram\_tadj@mail.ru}

\begin{abstract}
We assess a surface-density anisotropy diagnostic, $\Aani$, and its association with the slope $\alpha$ of a power-law galaxy profile. We reprocess 20-annulus profiles of 85 optically rich clusters on a canonical grid, $0\leq\Aani\leq1.92$, and an extended grid reaching 1.99. Twenty-four canonical fits and 12 extended fits reach their respective upper boundaries; 69 of 85 upper $\Delta\chi^2=1$ bounds remain open, and only four clusters have two-sided intervals. Among 40 preselected power-law fits, the observed $\alpha$--$\Aani$ Spearman coefficient is $0.616$. In a model-native forward-recovery experiment with 11 injected values, all 935 noiseless profiles recover the input within 0.02. Across 46,750 Poisson catalogs matched to observed count and background conditions, recovery instead has root-mean-square error (RMSE) $0.992$ and Spearman $\rho=0.093$; 40.0\% of fits reach the canonical upper boundary and 81.4\% have an open upper interval. Fixing both background and radial scale improves RMSE to $0.564$ in a 9,350-catalog sensitivity test. As a separate structural control, 4,250 Poisson catalogs generated from fitted power laws reproduce the observed slope association: the median mock coefficient for the 40-cluster subset is $0.608$ (95\% range $0.453$--$0.731$). Thus the diagnostic is numerically recoverable under its defining model without noise, but the available photometry does not reliably identify it for individual clusters. The power-law correlation can arise from shared projected profile shape and is not independent evidence for orbital anisotropy.
\end{abstract}


\section{INTRODUCTION}

The orbital distribution of galaxies in clusters contains information about accretion and dynamical evolution. In a spherical system, the velocity-anisotropy profile is conventionally written as

\begin{equation}
\beta(r)=1-\frac{\sigma_{\theta}^{2}(r)+\sigma_{\phi}^{2}(r)}{2\sigma_{r}^{2}(r)}.
\label{eq:beta}
\end{equation}

Inference of $\beta(r)$ from projected phase space is coupled to the mass profile and to tracer selection, producing the familiar mass--anisotropy degeneracy \citep{BinneyMamon1982,Merritt1987,LokasMamon2003,MamonLokas2005}. Modern projected-phase-space methods reduce this degeneracy by fitting the tracer density, mass profile, and orbital model together, sometimes with external lensing information, but their constraints remain conditional on equilibrium, geometry, membership, and parametric choices \citep{Mamon2013,AguirreTagliaferro2021,Pizzuti2021}.

The observational literature provides a broad but non-universal picture. Analyses of nearby and intermediate-redshift clusters have often found approximately isotropic inner orbits and a tendency toward greater radial anisotropy at larger radii, with differences among passive, star-forming, red, and blue tracers \citep{Biviano2004,BivianoPoggianti2009,Hwang2008,Wojtak2010,Capasso2019}. Comparable behavior has been reported to $z\gtrsim1$, although the uncertainties and population dependence remain important \citep{Biviano2013,Biviano2016,Biviano2021}. Recent cluster-by-cluster, stacked, and simulation studies reinforce two points: $\beta(r)$ is normally a radial function rather than a single scalar, and genuine cluster-to-cluster or tracer-to-tracer variance can be substantial \citep{Li2023,Ferrami2023,Barrena2024,ValkRembold2025,Abdullah2025,Pizzuti2025,Biviano2026,Maraboli2026}.

The method examined here is different. It assigns a constant anisotropy parameter by matching a family of stationary spherical density models to projected galaxy counts \citep{Kutlimuratov2024}. The parameter is physically defined within that model, but no individual line-of-sight velocities enter the fit. Its use therefore requires two separate validations: recovery under the defining model and external confirmation that real clusters satisfy the assumptions closely enough for the recovered parameter to trace orbital anisotropy.

We address the first requirement with a model-native forward experiment,
\[
\Aani^{\rm true}\longrightarrow D(x;\Aani^{\rm true})
\longrightarrow \Sigma(R;\Aani^{\rm true})
\longrightarrow n_i
\longrightarrow \Aani^{\rm fit}.
\]
We test numerical recovery without noise, recovery under the observed count and background conditions, sensitivity to nuisance parameters, and a power-law Poisson negative control. Our specific aim is to determine whether the surface-density diagnostic is identified by the available photometry and whether its measured association with the power-law slope exceeds the association expected from fitting both quantities to the same projected counts.

\section{SAMPLE AND METHODS}

\subsection{Cluster Sample and Annular Profiles}

The parent material contains 213 rich clusters selected from an SDSS-based catalog and cross-identified through the NASA/IPAC Extragalactic Database (NED) and VizieR. The working sample retains 85 systems with catalog optical richness $R_{\rm L}\geq100$, a visually identifiable central concentration, and $0.05<z<0.5$ \citep{Wen2012,Wen2015,Kutlimuratov2024}. These criteria define a targeted, high-contrast sample rather than a volume-complete population.

SDSS DR12 image cutouts of $600\times600$ pixels were divided into 20 equal-area annuli. With an inner radial scale of $r_1=40$ pixels, the outer radius of annulus $i$ is $r_i=r_1\sqrt{i}$ and every annulus has area $\Delta S=\pi r_1^2$. If $n_i$ galaxies are counted in annulus $i$, the adopted surface density and its Poisson approximation are

\begin{equation}
\Sigma_i=\frac{n_i}{\Delta S},\qquad
\sigma_{\Sigma,i}=\frac{\sqrt{n_i}}{\Delta S}.
\label{eq:counts}
\end{equation}

The fixed pixel scale is a reproducible observational coordinate, not a fixed physical aperture. Cluster centering choices and genuine breaks or curvature in projected galaxy profiles can modify the inferred inner slope and outer structure \citep{Rykoff2014,More2016,Baxter2017}. Visual centering, projection contamination, masking, and low-count departures from Gaussian errors are not included in the quoted parameter uncertainties.

\subsection{Surface-density Anisotropy Diagnostic}

Following \citet{Kutlimuratov2024}, the constant diagnostic is

\begin{equation}
\Aani=2-\frac{\sigma_t^2}{\sigma_r^2}
=2-\frac{\sigma_{\theta}^{2}+\sigma_{\phi}^{2}}{\sigma_r^2}
=2\beta,
\label{eq:aani}
\end{equation}

where $\sigma_t^2=\sigma_{\theta}^{2}+\sigma_{\phi}^{2}$. Thus $\Aani=0$ corresponds to isotropy in the model and $\Aani\rightarrow2$ to a radially dominated limit. This interpretation assumes a stationary, spherical, self-gravitating system with a spatially constant anisotropy and with the dispersion treatment used in the published implementation.

The underlying dimensionless density profile, $D(x)$, is obtained by combining the stationary radial force equation with Poisson's equation. In the notation of \citet{Kutlimuratov2024}, the constant-dispersion form is

\begin{equation}
\frac{d}{dx}\left(\frac{x^2}{D}\frac{dD}{dx}\right)
=-D x^2-\Aani.
\label{eq:densityode}
\end{equation}

For each trial $\Aani$, the numerical solution is projected along the line of sight and scaled in radius and normalization. Writing the resulting dimensionless projected family as $F_{\Aani}$, the comparison model is

\begin{equation}
\Sigma_{\rm mod}(R;\Aani,C,s,\Sigma_{\rm bg})
=C F_{\Aani}(R/s)+\Sigma_{\rm bg},
\label{eq:projectedmodel}
\end{equation}

where $C$, $s$, and the spatially constant background $\Sigma_{\rm bg}$ are fitted nuisance parameters. For a fixed $\Aani$ and $s$, the non-negative amplitude and background are obtained by bounded weighted least squares. The profile objective is

\begin{equation}
\chi^2(\Aani)=\sum_{i=1}^{20}
\frac{[\Sigma_i-\Sigma_{\rm mod}(R_i;\Aani,C,s)]^2}
{\sigma_{\Sigma,i}^{2}}.
\label{eq:chi2}
\end{equation}

\subsection{Likelihood Reprocessing and Boundary Definitions}

We independently recomputed the complete profile objective for the 85 annular data sets. The canonical grid contains $\Aani=0.00,0.04,\ldots,1.92$. An extended diagnostic grid uses a spacing of 0.01 and reaches 1.99, close to the mathematical radial limit without evaluating the singular endpoint. Radial scales are profiled on a logarithmic grid from 0.02 to 30 arcmin and locally interpolated around the best scale. For mock bins with zero counts, the variance safeguard is $\sigma_{\Sigma,i}=\sqrt{\max(n_i,1)}/\Delta S$.

We call a solution a \emph{canonical boundary fit} when the minimum on the canonical grid occurs at 1.92 and an \emph{extended boundary fit} when the fine-grid minimum occurs at 1.99. The $\Delta\chi^2=1$ set is evaluated from the complete fine-grid objective. An interval is upper-open if the objective does not cross $\chi^2_{\min}+1$ before 1.99 and two-sided only if both crossings occur inside the grid. These definitions replace the uncalibrated limits in the earlier transcribed output.

\subsection{Power-law Sample and Reliability Screen}

We fit the same annular profiles with a power law and constant background. The 40-object association subset was selected using finite fit and covariance values, formal uncertainties smaller than the magnitudes of the non-background parameters, positive admissible parameters, and solutions away from imposed numerical bounds. The screen is an operational reliability rule, not a formal identifiability theorem. Its row-level fits and screened subset are included in the accompanying structural-fit workbook and joined to the anisotropy table by cluster identifier. For

\begin{equation}
\Sigma_{\rm PL}(R)=A_{\rm PL}R^{-\alpha}+\Sigma_{\rm bg},
\label{eq:powerlaw}
\end{equation}

we compare the reprocessed extended-grid $\Aani$ with the amplitude $A_{\rm PL}$ and slope $\alpha$. Because the amplitude depends on the adopted radial unit and covaries with slope, the two relations are interpreted jointly.

\subsection{Model-native Forward-recovery Mocks}

The primary calibration follows equations~\ref{eq:densityode}--\ref{eq:chi2} in the forward direction. We inject
\begin{equation}
\begin{array}{rcl}
\Aani^{\rm true} & \in & \{0.10,0.30,0.50,0.70,0.90,1.10,\\
&&1.30,1.50,1.70,1.85,1.95\}.
\end{array}
\label{eq:injected}
\end{equation}
For each value, the density equation is solved independently with a higher numerical resolution than the recovery library and projected into the same 20 equal-area annuli. The radial scale is set to 7.7615 arcmin, an exact node of the recovery scale grid. For every cluster, the expected source plus background counts are normalized to its observed total count. The injected background is the cluster-specific value from the reprocessed fit, capped below 99.5\% of the aperture total.

This construction gives 935 noiseless profiles ($85\times11$). We additionally generate 50 independent Poisson realizations per cluster and injected value, giving 46,750 catalogs. Each catalog is refitted without knowledge of the input value on both the canonical and fine grids. We record bias, mean absolute error, root-mean-square error (RMSE), Spearman's $\rho$ between $\Aani^{\rm true}$ and $\Aani^{\rm fit}$, boundary frequencies, and membership of the injected value in the $\Delta\chi^2\leq1$ set.

To identify the source of any loss of information, we repeat the recovery on 10 realizations per cluster and injected value (9,350 catalogs per configuration) in four modes: free scale and free background, free scale and known background, known scale and free background, and both nuisance parameters known. This is a sensitivity experiment rather than a claim that either nuisance parameter is known for the observations.

\subsection{Power-law Poisson Negative Control}

A second mock experiment tests the power-law slope relation. For each of the 85 clusters, Poisson counts are drawn from its best-fitting $\Sigma_{\rm PL}(R)$ in equation~\ref{eq:powerlaw}. Fifty realizations per cluster give 4,250 catalogs. Both the generating power law and the $\Aani$ diagnostic are refitted to the same counts. The association is evaluated in the predefined 40-cluster subset for each of the 50 catalog realizations. These mocks contain no independently injected orbital information. If they reproduce the observed $\alpha$--$\Aani$ relation, shared profile shape and counting noise are sufficient to generate that relation. All 4,250 power-law and diagnostic refits converged; structural parameter reliability is evaluated separately.

\subsection{Statistical Summaries}

Pearson's $r$ summarizes linear association and Spearman's $\rho$ summarizes rank association. All reported $p$-values are two-sided and nominal. They do not propagate uncertainty in $\Aani$ or the structural fits, account for reliability-screen selection, or adjust for exploratory comparisons. Boundary ties weaken the usual continuous-sample approximation. For mock catalog-level correlations, the quoted interval is the 2.5th--97.5th percentile across 50 realizations. Cluster-bootstrap intervals resample the 85 observational conditions rather than treating 46,750 catalogs as independent clusters. The coefficients remain descriptive effect sizes, not population-level causal estimates.

\section{RESULTS}

\subsection{Reprocessed Distribution and Boundary Saturation}

\begin{figure*}[t]
\includegraphics[width=\textwidth]{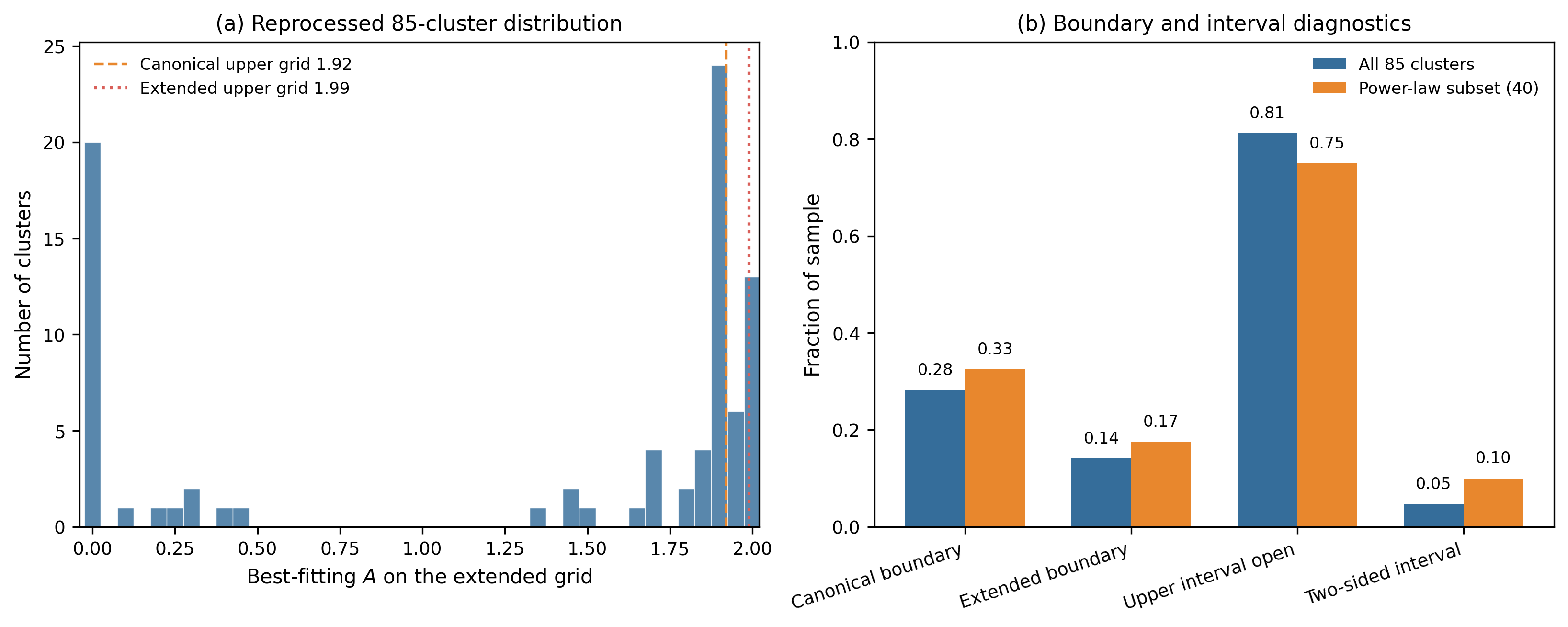}
\caption{Reprocessed surface-density diagnostic. Left: best-fitting values on the extended grid for 85 clusters; the canonical and extended upper grid values are marked separately. Right: fractions of the full sample and reliability-screened power-law subset with boundary minima or open/two-sided $\Delta\chi^2=1$ intervals. An extended-grid interior minimum can still have an open upper interval.}
\label{fig:distribution}
\end{figure*}

The independently recomputed likelihoods are strongly non-Gaussian and frequently open at one or both ends of the evaluated range. Table~\ref{tab:boundaries} and Figure~\ref{fig:distribution} separate a grid-end minimum from an interval that remains open. In the full sample, 24 of 85 canonical-grid minima occur at $\Aani=1.92$. Extending the search to 1.99 reduces the number of grid-end minima to 12, but 69 upper $\Delta\chi^2=1$ bounds remain open. Only four clusters have two-sided intervals.

\begin{deluxetable}{lcc}
\tablecaption{Boundary and Interval Diagnostics\label{tab:boundaries}}
\tablehead{
\colhead{Diagnostic} & \colhead{All 85} & \colhead{Power-law Subset}
}
\startdata
Canonical upper-bound minimum & 24 & 13 \\
Extended upper-bound minimum & 12 & 7 \\
Upper $\Delta\chi^2=1$ bound open & 69 & 30 \\
Two-sided $\Delta\chi^2=1$ interval & 4 & 4 \\
\enddata
\tablecomments{The canonical and extended upper grid values are 1.92 and 1.99, respectively. The power-law subset contains 40 clusters passing the structural reliability screen.}
\end{deluxetable}

The extended-grid values have mean 1.296, median 1.880, and sample standard deviation 0.851. Their distribution is concentrated near zero and near the radial limit, rather than resolving a smooth sequence of intermediate values. The median source fraction in the mock normalization is 0.228, so the background and radial scale can absorb much of the modest shape difference among trial values of $\Aani$.

\subsection{Power-law Structural Coupling}

\begin{figure*}[t]
\includegraphics[width=\textwidth]{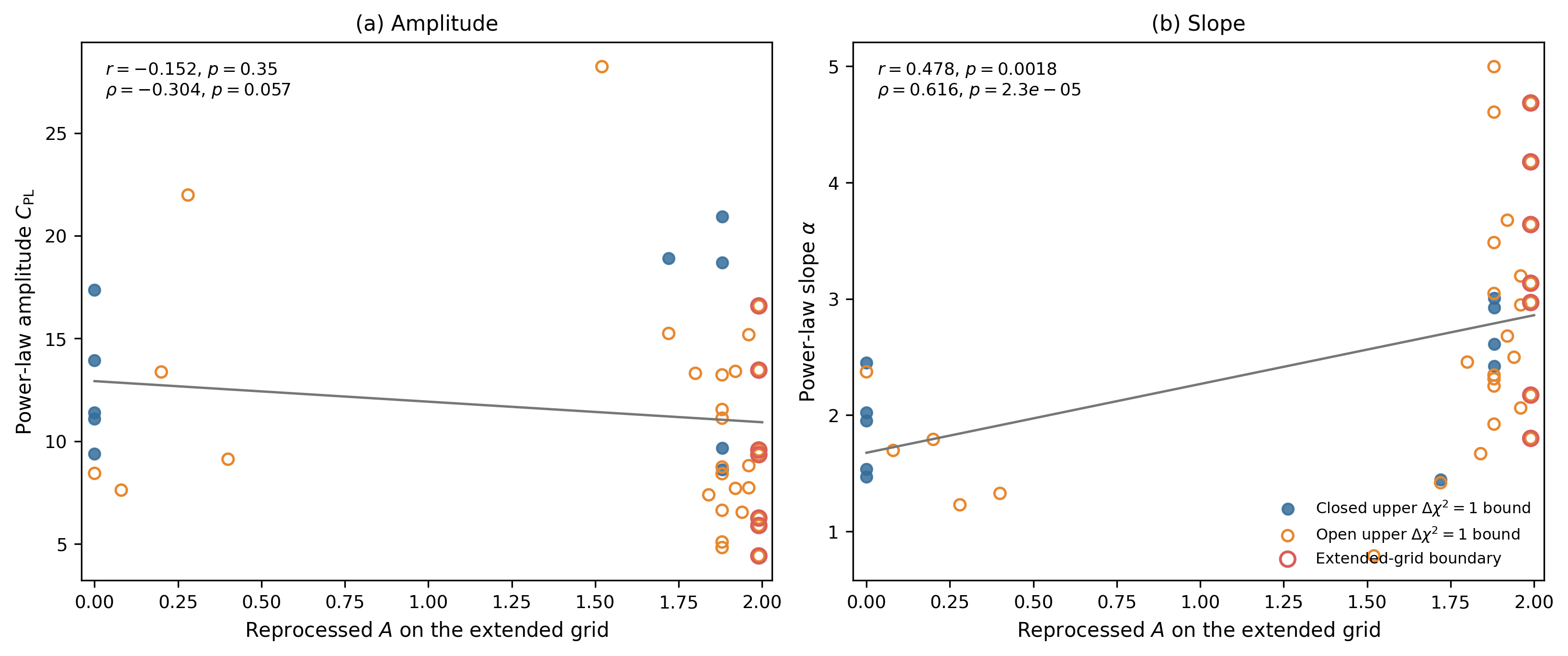}
\caption{Power-law amplitude (left) and slope (right) versus the reprocessed extended-grid diagnostic for the 40 clusters passing the structural reliability screen. Orange symbols identify open upper $\Delta\chi^2=1$ bounds, and red outlines mark extended-grid minima. The gray lines are ordinary least-squares fits. Both axes are derived from the same annular counts.}
\label{fig:correlations}
\end{figure*}

Table~\ref{tab:correlations} gives the two associations evaluated in the reliability-screened power-law subset. Figure~\ref{fig:correlations} shows the corresponding scatter plots and ordinary least-squares lines.

\begin{deluxetable*}{lccccc}
\tablecaption{Associations in the Reliability-screened Power-law Subset\label{tab:correlations}}
\tablehead{
\colhead{Relation} & \colhead{$N$} & \colhead{Pearson $r$} & \colhead{$p_r$} & \colhead{Spearman $\rho$} & \colhead{$p_\rho$}
}
\startdata
$\Aani$ versus $A_{\rm PL}$ & 40 & $-0.152$ & $0.348$ & $-0.304$ & $0.0569$ \\
$\Aani$ versus $\alpha$ & 40 & $0.478$ & $0.00181$ & $0.616$ & $2.33\times10^{-5}$ \\
\enddata
\tablecomments{$\Aani$ is the independently reprocessed extended-grid value. The $p$-values are two-sided and unadjusted; they do not include parameter uncertainty, screening uncertainty, or boundary censoring.}
\end{deluxetable*}

The reprocessed amplitude relation is weak, whereas the slope relation remains positive. After excluding the seven extended-grid boundary minima, the slope association remains present among 33 clusters (Pearson $r=0.463$, $p=0.0067$; Spearman $\rho=0.628$, $p=9.1\times10^{-5}$). This deletion is not a correction for censoring, but it shows that the sign is not produced solely by tied boundary points.

\subsection{Model-native Recovery}

The noiseless test verifies numerical self-consistency. All 935 profiles recover $\Aani^{\rm true}$ within 0.02; the global fine-grid RMSE is 0.0037. Adding Poisson noise at the observed count and background levels changes the result qualitatively. Across 46,750 catalogs, the fine-grid RMSE is 0.992 and the rank association between input and output is only $\rho=0.093$. The 95\% cluster-bootstrap interval for RMSE is 0.980--1.003. The canonical boundary fraction is 0.400 (bootstrap interval 0.384--0.417), 81.4\% of upper $\Delta\chi^2=1$ bounds are open, and only 6.1\% of intervals are two-sided.

\begin{figure*}[t]
\includegraphics[width=\textwidth]{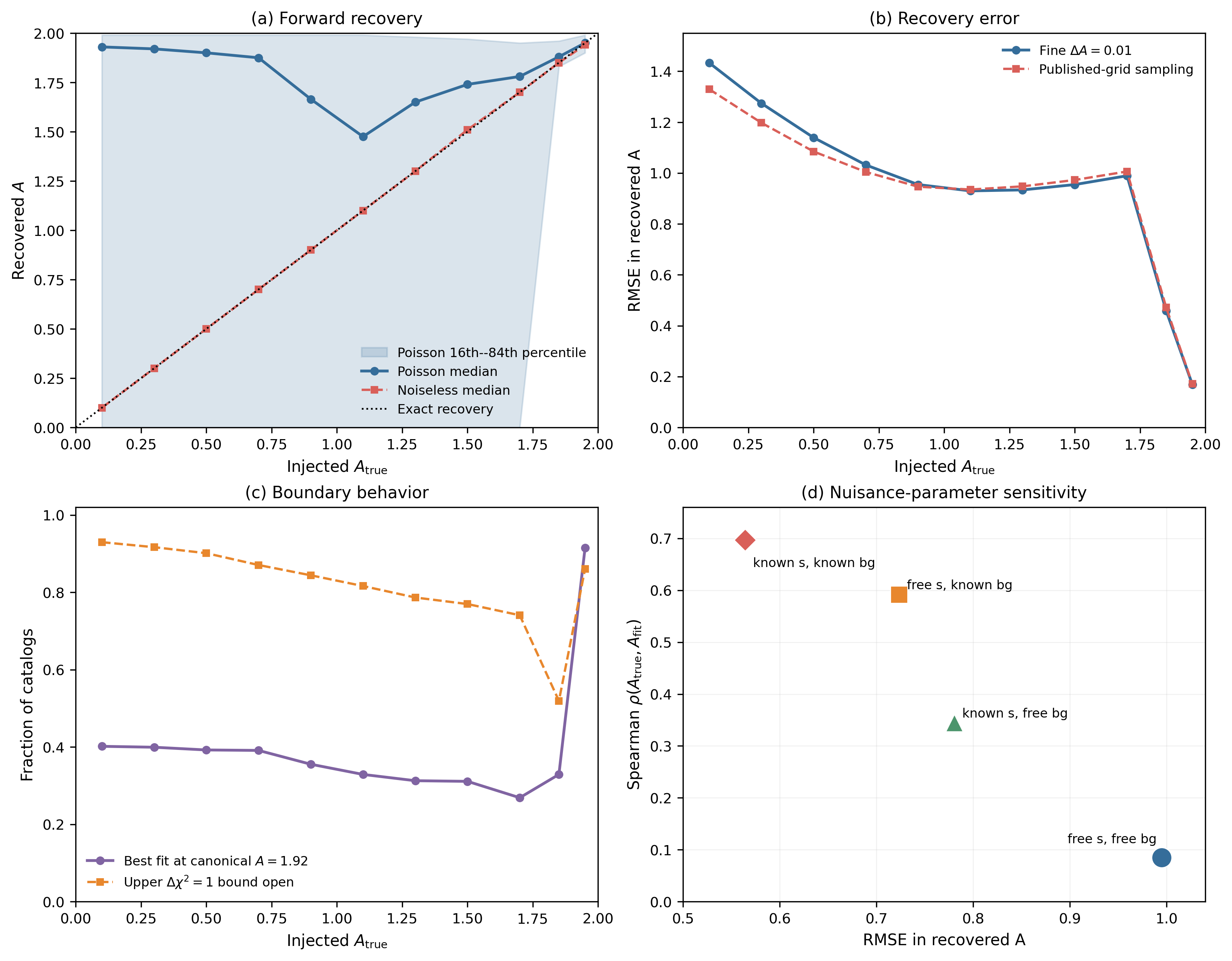}
\caption{Model-native forward-recovery results. (a) Median recovered diagnostic and 16th--84th percentile range for the 46,750 Poisson catalogs, compared with noiseless recovery and the one-to-one line. (b) RMSE on the fine and published-grid samplings. (c) Canonical boundary and open-upper-interval fractions. (d) Sensitivity to fixing the injected radial scale $s$ and background. Point size in panel (d) increases with the canonical boundary fraction.}
\label{fig:native}
\end{figure*}

Recovery depends strongly on the injected value. For $\Aani^{\rm true}=0.10$, the median recovered value is 1.93 and the RMSE is 1.43; for $\Aani^{\rm true}=1.95$, the median is 1.95 and the RMSE is 0.169. Thus a high fitted value is not uniquely generated by a high input value: low and intermediate inputs can also be driven to the upper part of the grid under weak source contrast.

Table~\ref{tab:nuisance} isolates the nuisance degeneracy. Knowing the background improves $\rho$ from 0.085 to 0.592 and lowers RMSE from 0.995 to 0.723. Knowing both background and scale gives $\rho=0.697$ and RMSE 0.564. The remaining error shows that Poisson sampling and residual shape similarity still limit recovery even under favorable nuisance information.

\begin{deluxetable*}{lcccccc}
\tablecaption{Model-native Mock Recovery\label{tab:nuisance}}
\tablehead{
\colhead{Configuration} & \colhead{$N_{\rm cat}$} & \colhead{RMSE} & \colhead{Spearman $\rho$} & \colhead{$\Delta\chi^2\leq1$ Coverage} & \colhead{Canonical Boundary} & \colhead{Fine Boundary}
}
\startdata
Primary, free $s$ and background & 46,750 & 0.992 & 0.093 & 0.833 & 0.400 & 0.200 \\
Sensitivity, free $s$ and background & 9,350 & 0.995 & 0.085 & 0.837 & 0.399 & 0.197 \\
Free $s$, known background & 9,350 & 0.723 & 0.592 & 0.836 & 0.197 & 0.105 \\
Known $s$, free background & 9,350 & 0.780 & 0.344 & 0.771 & 0.230 & 0.142 \\
Known $s$ and background & 9,350 & 0.564 & 0.697 & 0.731 & 0.129 & 0.084 \\
\enddata
\tablecomments{The sensitivity experiment uses the first 10 realizations for every cluster--input combination. The canonical and fine upper boundaries are $\Aani=1.92$ and 1.99, respectively.}
\end{deluxetable*}

\subsection{Power-law Negative Control}

\begin{figure*}[t]
\includegraphics[width=\textwidth]{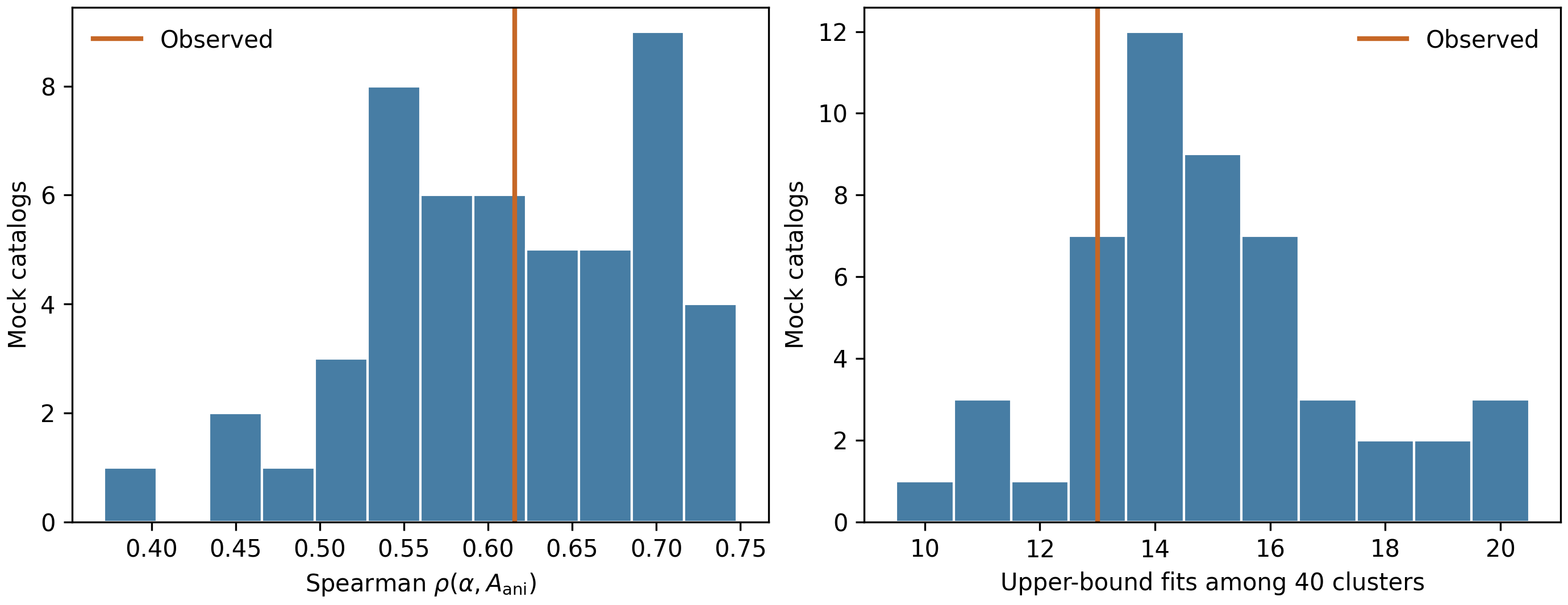}
\caption{Power-law Poisson negative control for the 40-cluster subset. Left: distribution of the 50 mock-catalog Spearman coefficients between fitted slope and diagnostic, with the observed coefficient marked by a vertical line. Right: mock-catalog distribution of the number of canonical upper-bound fits, with the observed count marked.}
\label{fig:plmock}
\end{figure*}

Figure~\ref{fig:plmock} and Table~\ref{tab:plmock} compare the observed slope association and boundary diagnostics in the 40-cluster subset with 50 power-law Poisson catalog realizations. The observed $\rho(\alpha,\Aani)=0.616$ is close to the mock median 0.608 and inside the mock 95\% range $[0.453,0.731]$. The observed number of canonical upper-bound fits is 13, versus a mock median of 15, while the observed number with an open upper interval is 30, identical to the mock median. These are catalog-level intervals, not uncertainty intervals on individual clusters.

\begin{deluxetable*}{lcccc}
\tablecaption{Power-law Poisson Negative Control for the 40-cluster Subset\label{tab:plmock}}
\tablehead{
\colhead{Diagnostic} & \colhead{$N$} & \colhead{Observed} & \colhead{Mock Median} & \colhead{Mock 95\% Range}
}
\startdata
Slope--$\Aani$ Spearman $\rho$ & 40 & 0.616 & 0.608 & [0.453, 0.731] \\
Canonical upper-bound fits & 40 & 13 & 15 & [11, 20] \\
Upper-open $\Delta\chi^2=1$ intervals & 40 & 30 & 30 & [25.2, 33.0] \\
Two-sided $\Delta\chi^2=1$ intervals & 40 & 4 & 5 & [2.2, 8.0] \\
\enddata
\tablecomments{Every realization refits the power law and $\Aani$ to the same Poisson counts. The ranges are 2.5th--97.5th percentiles across 50 catalogs, with 40 predefined clusters per catalog. The interval counts use the extended $\Aani$ grid.}
\end{deluxetable*}

The power-law negative control reproduces both the observed slope association and boundary saturation without independently injecting orbital information. It establishes a sufficient structural explanation for the measured correlation, while allowing that additional physical effects may also be present.

\section{DISCUSSION}

\subsection{What the Forward Test Establishes}

The noiseless result shows that equations~\ref{eq:densityode}--\ref{eq:chi2} and their numerical implementation contain recoverable information about $\Aani$ when the assumed profile is known exactly. This is a positive validation of the model map; it distinguishes the parameter from an arbitrary empirical label attached after fitting.

The Poisson experiment answers a different question. Under the observed total counts and fitted background fractions, the profile likelihood is usually too broad or multimodal to identify the injected value. The poor recovery is not repaired by refining the $\Aani$ grid: fine and published-grid RMSE values are nearly identical over most of the input range. It follows that a boundary result in the observations cannot be interpreted as a precise measurement close to the radial limit.

\subsection{Interpretation of the Structural Coupling}

The empirical slope relation and the power-law negative control use summaries extracted from the same projected counts. A steeper power-law profile is assigned a larger $\Aani$, but the nearly identical observed and mock coefficients show that this relation can arise from the shared profile shape. The negative control quantitatively demonstrates that independent orbital information is not required to generate a correlation of the observed magnitude.

This does not make $\Aani$ physically meaningless. Under the stationary spherical constant-anisotropy equations, it remains related to the velocity-dispersion ratio by equation~\ref{eq:aani}. The limitation is inferential: several combinations of $\Aani$, radial scale, normalization, and background generate nearly indistinguishable projected profiles over 20 annuli. Centering, radial coverage, and genuine profile curvature can add further degeneracy because photometric centers are probabilistic and cluster profiles can contain outer transitions \citep{Rykoff2014,More2016,Baxter2017}.

\subsection{Nuisance Information and External Dynamical Tests}

The sensitivity experiment identifies two practical improvements. An independently measured background removes a large part of the low-$\Aani$ ambiguity, while an externally constrained radial scale provides additional information. Even with both fixed, however, the RMSE remains 0.564 and the $\Delta\chi^2\leq1$ coverage is 0.731. A useful future estimator would therefore require joint calibration of nuisance parameters and a likelihood construction tested for coverage, not only a finer parameter grid.

The model-native test is internal: it establishes behavior when real clusters obey the defining density equation. It does not establish that their orbital anisotropy follows the fitted scalar. We recommend an independent projected phase-space analysis that estimates mass and $\beta(r)$ jointly for clusters with adequate member redshifts \citep{Mamon2013}. The relevant comparison is a pre-defined, tracer- and aperture-matched average of $\beta(r)$, because kinematic studies generally find a radial profile that is close to isotropic in cluster cores and more radial toward the virial region \citep{Capasso2019,Biviano2021,Li2023,ValkRembold2025}. Simulations further predict dependence on mass, redshift, interlopers, and viewing direction \citep{AguirreTagliaferro2021,Abdullah2025}. Spectroscopic or phase-space validation remains necessary before an individual photometric value is interpreted as an orbital measurement.

\subsection{Limitations}

The sample is selected for high richness and visible central concentration and is not representative of all clusters; apparent concentration alone does not exclude kinematic substructure or merger activity \citep{Pinkney1996}. The diagnostic assumes spherical stationarity, self-gravity, constant anisotropy, and the particular density family in equation~\ref{eq:densityode}. It does not use velocities. The forward mocks reproduce observed counts and fitted background fractions, but include only independent Poisson noise. They do not include correlated background uncertainty, masking, miscentering, substructure, line-of-sight projections, magnitude-dependent completeness, or departures from the assumed density equation. The nuisance sensitivity uses 10 rather than 50 realizations per cluster--input combination. Structural and diagnostic fits share data, and the structural reliability screen changes the analyzed subset. These restrictions prevent individual-cluster orbital classifications or a calibrated population distribution.

\section{CONCLUSIONS}

The model-native test resolves the main methodological question. In the absence of noise, the projected-density implementation recovers the injected anisotropy parameter to within 0.02 for all 935 tests. The parameter is therefore well defined and numerically recoverable under the stationary spherical model.

The observational information content is insufficient for precise cluster-by-cluster recovery. In 46,750 Poisson catalogs matched to the 85 observed count and background conditions, RMSE is 0.992, $\rho(\Aani^{\rm true},\Aani^{\rm fit})=0.093$, 40.0\% of fits reach the canonical upper boundary, and 81.4\% have an open upper interval. Fixing background and radial scale improves recovery but does not eliminate the error.

The 4,250 power-law Poisson catalogs reproduce the observed slope--$\Aani$ association in the selected 40 clusters. The measured association can therefore arise from coupling within the projected-density pipeline and cannot by itself establish an orbital relation.

The defensible interpretation is conditional: projected counts constrain $\Aani$ only under the adopted density model and only when nuisance information and signal-to-background contrast are adequate. We interpret boundary values from the present 20-annulus photometry as unresolved or one-sided constraints. External spectroscopy remains necessary before $\Aani$ is used as an orbital-anisotropy estimator.

\begin{acknowledgments}
We used Sloan Digital Sky Survey imaging, the NASA/IPAC Extragalactic Database, and the VizieR catalog service. We used OpenAI's ChatGPT under our supervision to assist with code translation and implementation, mock-catalog calculations, numerical checks, figure preparation, and English-language editing. We tested the computational analysis by independently reprocessing the 85 observed profiles, checking noiseless recovery for the injected model values, and recalculating the principal summary statistics from the archived row-level mock outputs. The accompanying code, inputs, random seeds, and outputs permit additional checks. We did not use AI to generate or alter observational data. We take responsibility for the methods, inputs, mock outputs, manuscript, and scientific conclusions.
\end{acknowledgments}

\begin{contribution}
S.S.K. assembled the sample, performed the original analysis, and drafted the manuscript. I.U.T. supervised the underlying dissertation research and contributed to the scientific interpretation.
\end{contribution}

\section*{Data Availability}

SDSS imaging is publicly available through the SDSS Science Archive Server. Catalog quantities are available from the sources cited in Section~2. The arXiv ancillary file \texttt{OJA\_reproducibility\_bundle.zip} contains the reprocessed 85-cluster diagnostic table, the 46,750-row model-native mock table, the 4,250-row power-law mock table, summary tables, 85 annular count files, random seeds, and Python scripts for forward generation, recovery, sensitivity tests, and figures. Large objective-profile arrays can be regenerated from those scripts; we can also provide them on request.

\section*{Conflict of Interest}

We declare no conflict of interest.

\end{document}